\documentclass[aps,showpacs,preprint,amsmath,amssymb]{revtex4}
\usepackage{amsmath}
\usepackage{epsfig}

\newcommand{\bra}{\langle}
\newcommand{\ket}{\rangle}
\newcommand{\threej}[6]{
     \left( \begin{array}{ccc}
              #1 & #2 & #3 \\
              #4 & #5 & #6 
            \end{array}  \right) } 

\newcommand{\sixj}[6]{
     \left\{ \begin{array}{ccc}
              #1 & #2 & #3 \\
              #4 & #5 & #6 
            \end{array}  \right\} } 
\newcommand{\ninej}[9]{
     \left\{ \begin{array}{ccc}
              #1 & #2 & #3 \\
              #4 & #5 & #6 \\
              #7 & #8 & #9 
            \end{array}  \right\} } 

\begin{document}
\title{Incorporation of three--nucleon force in the effective interaction
hyperspherical harmonic approach}

\author{Nir Barnea$^1$,
  Victor D. Efros$^{2}$,
  Winfried Leidemann$^{3}$ 
  and Giuseppina Orlandini$^{3}$ 
}
\affiliation{$^{1}$ The Racah Institute of Physics - The Hebrew University, 
91904, Jerusalem, Israel \\ $^{2}$Russian Research Centre "Kurchatov Institute"
- 123182 Moscow, Russia \\ $^{3}$Dipartimento di Fisica - Universit\`a di Trento, 
and Istituto Nazionale di Fisica Nucleare - Gruppo Collegato di Trento, I-38050 
Povo, Italy\\
}

\date{\today}

\begin{abstract}
It is shown how a bare three--nucleon force is incorporated into the formalism
of the effective interaction approach for hyperspherical harmonics. As a
practical example we calculate the ground state properties of $^3$H and $^3$He 
using the Argonne V18 nucleon--nucleon potential and the Urbana IX 
three--nucleon force. A very good convergence of binding energies and 
matter radii is obtained. We also find a very good agreement of our results 
compared to other high precision calculations. 

\end{abstract}

\bigskip

\maketitle
\section{Introduction}

It is well known that the binding energies of the three--body nuclei cannot be 
obtained with NN forces only and thus also three--nucleon (3N) forces have to 
be included. This makes 
realistic calculations for systems with more than two nucleons more difficult. 
Nonetheless such calculations have been carried out in three-- and 
four--nucleon systems by various groups using different theoretical methods: 
Green Function Monte Carlo \cite{GFMC} (also for $A > 4$) as well as 
hyperspherical harmonics (HH) expansions and Faddeev techniques (see e.g. 
\cite{ELOT,ELOT2,BoPi,Nogga}). It would be very desirable 
to extend such an exact treatment of 3N forces to methods based on 
effective interaction approaches, since they are very suitable to treat also 
systems with $A > 4$. In case of the effective interaction approach within
the No Core Shell Model (NCSM) such calculations have already been performed. 
Rather good results have been obtained \cite{NCSM}, though without reaching 
a convergence as excellent as in other methods. For the effective 
interaction hyperspherical harmonics (EIHH) technique \cite{EIHH1} an 
inclusion of bare 3N forces had not yet been accomplished. In fact the aim of 
the present paper is to show how such a 3N force is incorporated in the EIHH 
formalism. A further aim is to check whether a comparably good convergence as 
with methods, which do not use effective interaction approaches, can be reached.
To this end results of a calculation of three--nucleon groundstate properties 
using the AV18 NN potential \cite{AV18} and the UrbIX 3N force \cite{UrbIX} are
presented.

The paper is organized as follows. In section II the EIHH method is briefly 
outlined, while the incorporation of the 3N force in the EIHH method is 
described in section III. Results for the three--nucleon groundstate properties
are discussed in section IV. Details for the calculation of the 3N force
matrix elements are given in the Appendix.

\section{Brief outline of the EIHH method}

To apply an effective interaction method to an $A$--body Hamiltonian of the 
form 
\begin{equation}\label{1}
  H^{[A]} = H_0+V
\end{equation}
one divides the Hilbert space of $H^{[A]}$ into a model space P and a residual 
space Q with eigenprojectors $P$ and $Q$ of $H_0$. The Hamiltonian $H^{[A]}$ 
is then replaced by an effective model--space Hamiltonian
\begin{equation}\label{2}
  H^{[A]eff} = PH_0 P + P V^{[A]eff} P \;
\end{equation}
that by construction has the same energy levels as the low--lying states of 
$H^{[A]}$. To find $V^{[A]eff}$, however, is as difficult as seeking the 
full--space solutions. In the EIHH method, as in the NCSM method, one 
approximates $V^{[A]eff}$ in such a way that it coincides with $V$ for 
P$\longrightarrow 1$, so that an enlargement of P leads to a convergence of 
the eigenenergies to the {\it true} values.

Considering a Hamiltonian that contains besides the kinetic energy $T$ 
only a
two--body interaction $v^{[2]}_{ij}$,
\begin{equation}\label{3}
  H^{[A]} = T +  \sum_{i<j}^{A} v^{[2]}_{ij} \,,
\end{equation}
one can rewrite $H$ in the hyperspherical formalism as
\begin{equation}\label{H^[A]}
  H^{[A]} = T_\rho+T_K(\rho) + V^{[A]}(\rho,\Omega_A)\,,
\end{equation}
where 
\begin{equation}\label{4}
  V^{[A]}(\rho,\Omega_A)\equiv\sum_{i<j}^{A} v^{[2]}_{ij}\,,\,\,\,\,\,
  T_\rho=- \frac{1}{2m}\Delta_{\rho}\,,\,\,\,\,\,
  T_K(\rho)=\frac{1}{2 m} \frac{\hat{K}_A^2}{\rho^2} 
\end{equation}
are the bare two--body potential and the hyperradial/hypercentrifugal 
kinetic energies, respectively, $m$ is the particle mass, $\hat{K}_A$ is the 
hyperspherical grand angular momentum operator and $\Omega_A$ are the 
$(3A-4)$--dimensional hyperangles, while $\Delta_{\rho}$ denotes the Laplace 
operator with respect to the hyperradial coordinate
\begin{equation}\label{5}
   \rho = \sqrt{\sum_{j=1}^{N} \vec{\eta}_j^{\,2}} \,\,\,\,\,\,\,\,\, {\rm with}
\,\,\,\,\,\,\,\,\,  \vec{\eta}_i = \sqrt{\frac{A-i}{A+1-i}}
  \left(\vec{r}_i - \frac{1}{A-i}\sum_{j=i+1}^A \vec{r}_j \right) \;.
\end{equation}
Here $\vec\eta_i$ are the $N=A-1$ Jacobi vectors in the reversed order and 
$\vec{r}_j$  are the position vectors of $A$ particles. In the EIHH method one 
takes   the so called adiabatic Hamiltonian as a  starting point,
\begin{equation}\label{6}  
  {\cal H}^{[A]}(\rho)\equiv T_K(\rho)  +  V^{[A]}(\rho,\Omega_A) \,,
\end{equation} 
and the unperturbed Hamiltonian $H_0$ is chosen to be $T_K(\rho)$, which has 
the HH basis functions ${\cal Y}_{[K_A]}$ as eigenfunctions, where 
$[K_A]$ stands for the appropriate set of quantum numbers 
(see \cite{EIHH1}). The P--space 
is defined as the complete set of HH basis functions with generalized 
angular momentum quantum number $K_A \leq K_P$ and correspondingly the Q--space 
with $K_A > K_P$. 

For each hyperradius $\rho$ an effective adiabatic Hamiltonian is constructed
\begin{equation}\label{7}
{\mathcal H}^{[A]eff}(\rho,\Omega_A)= P T_K(\rho) P  +  
                 P\,V^{[A]eff}(\rho,\Omega_A)\,P\,,
\end{equation}
where $V^{[A]eff}$ is approximated by a sum of {\it two--body} terms 
\begin{equation}\label{8}
  V^{[A]eff}\simeq \,\,\sum_{i<j}^A v^{[2]eff}_{i,j}\,.
\end{equation}
Due to the use of antisymmetric wave functions one needs to calculate  
$v^{[2]eff}_{i,j}$ only relative to one pair, since
\begin{equation}\label{9}   
  \langle\,V^{[A]eff}\, \rangle \simeq \langle\, \sum_{i<j}^A 
  v^{[2]eff}_{i,j}\,\rangle=\frac{A(A-1)}{2}
  \langle\, v^{[2]eff}_{A,(A-1)}\,\rangle \;.
\end{equation}

For the determination of $v^{[2]eff}_{A,(A-1)}$ one takes a $\rho$--dependent 
{\it quasi two--body} adiabatic Hamiltonian
\begin{equation}\label{10}
  {\mathcal H}^{[2]}(\rho\, ; \theta_N,\hat{\eta}_N) = 
  T_K(\rho) + v^{[2]}_{A,(A-1)}
  (\sqrt{2}\rho \sin \theta_N\cdot \hat{\eta}_N) \,,
\end{equation}
where the unit vector $\hat{\eta}_N$ and the hyperangle $\theta_N$ are 
defined by
\begin{equation}\label{eta_N}
 \hat{\eta}_N =  \frac{\vec\eta_N}{\eta_N} \,,\,\,\,\,\,
 \vec\eta_N = \sqrt{\frac{1}{2}}(\vec{r}_{A-1}-\vec r_A) \,,\,\,\,\,\,
 \eta_N = \rho \sin \theta_N \;.
\end{equation}
The Hamiltonian of Eq. (\ref{10}) is diagonalized on the $A$--body HH basis,
which then allows to apply the Lee--Suzuki \cite{LS} similarity 
transformation to ${\mathcal H}^{[2]}(\rho\,;\theta_N, \hat{\eta}_N)$. This
finally leads to an expression for a $\rho$--dependent two--body effective
interaction $v^{[2]eff}(\rho\,;\theta_N,\hat{\eta}_N)$ (for details see 
\cite{EIHH1}), and thus the $A$--body problem can be solved in the P--space with 
the following effective Hamiltonian 
\begin{equation}\label{HAeff}
H^{[A]eff}=T_\rho+{\mathcal H}^{[A]eff} \simeq 
                            T_\rho+T_K+\sum_{i<j}v^{[2]eff}_{ij} \,.
\end{equation}
One repeats the procedure enlarging the P--space up to a convergence 
of the low--lying energies of an $A$--body system. As shown in \cite{EIHH2} 
the EIHH method leads to a very fast convergence for the binding energies of 
few--nucleon systems with realistic NN potential models.

In \cite{EIHH3} we could show that the convergence is even further improved
if one goes beyond the adiabatic approximation taking into account also an
effective hyperradial kinetic energy $\Delta T^{eff}_\rho$. This leads to
the following nonadiabatic two--body effective interaction 
\begin{equation}\label{24}
  \tilde{v}^{[2]eff}_{A,A-1} 
      = {\mathcal H}^{[2]eff} - T_K(\rho)-T_\rho
      = v^{[2]eff}_{A,A-1} + \Delta T^{eff}_\rho\;.
\end{equation}
For explicit expressions of $\Delta T^{eff}_\rho$ we refer to \cite{EIHH3}.

\section{Incorporation of three--nucleon force}

The inclusion of a bare three--body interaction $v^{[3]}_{ijk}$ leads to 
the following HH $A$--body Hamiltonian 
\begin{equation}\label{H_A}
  H^{[A]} = T_\rho + T_K(\rho) + \sum_{i<j}^{A} v^{[2]}_{ij} + \sum_{i<j<k}^{A} 
v^{[3]}_{ijk} \,.
\end{equation}
While for the two--body interaction one needs to consider only the 
hyperspherical coordinates connected to the $A$-$(A-1)$ 
pair explicitly, for the three--body interaction one has to take 
into account additional coordinates. We were confronted with the same problem
when deriving in \cite{EIHH3} a three--body effective interaction from a
bare two--nucleon potential. Thus we can proceed here in the same manner.

In order to focus on the {\it interacting} three--body subsystem we transform 
the $(3A-4)$ hyperangular coordinates 
$\Omega_A =(\theta_2,\ldots,\theta_{A-1},\hat\eta_1,\ldots ,\hat \eta_{A-1})$
into a new set of hyperangles 
\begin{equation}\label{32}
   \Omega_{3,A-3} =(\theta_2,\ldots,\theta_{A-3},\Theta^{3,A-3},
  \theta^3_{int},\hat\eta_1,\ldots ,\hat\eta_{A-1})\,.
\end{equation}
The new hyperangles reflect the splitting of the $A$--body system into 3-- and 
$(A-3)$--body subsystems. The hyperangles  $\Omega_{3,A-3}$ can be written as
\begin{equation}\label{33}  
 \Omega_{3,A-3} = (\Theta^{[3,A-3]},\Omega^{[3]}_{int},\Omega^{[A-3]}_{res})
\end{equation}
with the hyperangles of the {\it  interacting} three--body subsystem 
$(\Omega^{[3]}_{int})$ and those of the residual system 
$(\Omega^{[A-3]}_{res})$ defined as
\begin{equation}\label{35}  
 \Omega^{[3]}_{int}   = (\theta^{[3]}_{int}, \hat\eta_{A-2},\hat\eta_{A-1}) \,,
 \,\,\,\,\,\,\,\,\,
      \Omega^{[A-3]}_{res} = (\theta_2,\ldots,\theta_{A-3},
                     ,\hat\eta_1,\ldots ,\hat\eta_{A-3}) \,.
\end{equation} 
The two new angles $\Theta^{[3,A-3]}$ and $\theta^{[3]}_{int}$, replacing 
$\theta_{A-2}$ and $\theta_{A-1}$, are given by the relations
\begin{equation}\label{36}
     \rho^{[3]}_{int} \equiv\sqrt{\eta_{A-1}^2+\eta_{A-2}^2} 
      = \rho \,\sin \Theta^{[3,A-3]} \,,\,\,\,\,\,\,\,\,\,\,
     \rho_{A-3}  =  \rho \,\cos \Theta^{[3,A-3]} 
\end{equation}
and
\begin{equation}\label{37}
     \eta_{A-1}  = \rho^{[3]}_{int}  \,\sin \theta^{[3]}_{int} \,,\,\,\,\,\,\,
     \,\,\,\, \eta_{A-2}  = \rho^{[3]}_{int}  \,\cos \theta^{[3]}_{int} \,.
\end{equation}
The new coordinates $(\rho^{[3]}_{int}, \Omega^{[3]}_{int})$ form a complete 
set for the three--body problem and thus matrix elements of 3N forces can
be calculated for such a system. 

The calculation  proceeds as follows. In a first step one solves the 
$A$--body Schr\"odinger equation with the Hamiltonian of Eq.~(\ref{10}) and 
constructs the nonadiabatic two--body effective interaction 
$\tilde{v}^{[2]eff}$. As next step one has two alternatives. One of
them is to solve the Schr\"odinger equation with the Hamiltonian 
\begin{equation}\label{HA}
H^{[A]eff} \simeq T_\rho+T_K+ \sum_{i<j}\tilde v^{[2]eff}_{ij}
         + \sum_{i<j<k} v^{[3]}_{ijk} \,.
\end{equation} 
This is the effective Hamiltonian, which is used in the three-nucleon 
calculations presented in section IV. 

The other alternative (meaningful for $A>3$ only) is to solve first
the adiabatic {\it quasi three--body} Hamiltonian
\begin{eqnarray}\label{H3}
  {\mathcal H}^{[3]}(\rho) & = &
      T_K(\rho) + v^{[2]eff}_{A,(A-1)}+ v^{[2]eff}_{(A-1),(A-2)} + 
        v^{[2]eff}_{(A-2),A} + v^{[3]}_{(A-2),(A-1),A}
          \cr & \equiv &
          T_K(\rho) + V^{[3]}(\rho^3_{int},\Omega^3_{int}) \,.
\end{eqnarray}
The so defined $V^{[3]}$ contains all three effective pair interactions 
$v^{[2]eff}_{ij}$ and different from \cite{EIHH3} also the bare 3N interaction 
$v^{[3]}_{ijk}$. Then one can proceed further in complete analogy to 
\cite{EIHH3}, i.e. along the same lines as for the construction of 
$\tilde{v}^{[2]eff}$ applying the Lee--Suzuki similarity transformation in 
order to obtain $\tilde{v}^{[3]eff}$. Finally one solves the 
$A$--body Schr\"odinger equation 
\begin{equation}
H^{[A]eff}=T_\rho+{\mathcal H}^{[A]eff} \simeq 
                  T_\rho+T_K+\sum_{i<j<k}\tilde v^{[3]eff}_{ijk}
\end{equation} 
taking into account that
\begin{equation}
  \langle\,\tilde V^{[A]eff}\, \rangle \simeq \langle\, \sum_{i<j<k}^A 
        \tilde v^{[3]eff}_{i,j,k}\,\rangle=\frac{A(A-1)(A-2)}{6}
        \langle\, \tilde v^{[3]eff}(A,A-1,A-2)\,\rangle \,.
\end{equation}

\section{Discussion of results}

In order to check the convergence behavior of the EIHH method in presence of 
3N forces we choose as test cases the ground states of $^3$H and $^3$He. We use 
the AV18 NN potential \cite{AV18} including the electromagnetic corrections of 
the NN force, but neglecting the isospin mixing. As 3N force we take the UrbIX 
model \cite{UrbIX}. The calculation of the 3N force matrix elements is 
described in the Appendix.

In Fig.~1 we show the binding energy of the triton as function of the 
grand angular momentum quantum number $K$. One observes a very good convergence 
pattern. In fact already with $K=8$ one has only a difference of about 50 keV
with the value of $K=20$. This difference is further reduced to about 20 keV
with $K=10$, while with $K=12$ one has essentially the final result. In Fig.~2
we show the corresponding results for $^3$He. One finds a very similar 
picture as in Fig.~1, i.e. again an excellent convergence.
In Fig.~3 we illustrate the results for the matter radii of $^3$H and $^3$He. 
Also in these cases a very good convergence is evident with almost
identical convergence patterns for $^3$H and $^3$He.

In Table~I we list our results for the triton binding energy and the 
probabilities for the different angular momentum components together with 
those of Table~I in \cite{ELOT2}, where the inelastic longitudinal electron 
scattering form factors of the three--nucleon systems were calculated
(for this comparison the electromagnetic part of AV18 is not included).
There is a good agreement for the various probabilities and a small difference 
of about 40 keV for the binding energy.

With an inclusion of the electromagnetic interaction of AV18 we obtain the 
results given in Table~II, which are compared to those of \cite{BoPi}.
The comparison of the various results reveals an 
excellent agreement for the various ground state properties obtained in the
different calculations, e.g., the maximal difference for the binding 
energy is 6 keV for both three--body nuclei.  
Thus we may conclude that 3N forces can be included in the EIHH formalism
with high precision. This is an important finding opening up the way to 
include 3N forces also for systems with $A>3$ in this approach.
 
\renewcommand{\theequation}{A.\arabic{equation}}
\setcounter{equation}{0}
\section*{Appendix: Multipole expansion and matrix elements of 3N forces}

We consider the Urbana \cite{UrbIX} and Tucson-Melbourne \cite{TM} versions of 
3N forces. To calculate their matrix elements (ME) we first represent the 3N 
force operators as expansions over irreducible space tensors. We shall list
these expansions in two different forms.

The  Urbana or Tucson-Melbourne 3N forces $v_{123}^{[3]}$ are given by a sum 
$\sum_{cyc} V_{ijk}$ taken over cyclic permutations of particles 1,2,3. Each 
of the three terms in the sum gives the same contribution to ME between states 
antisymmetric with respect to nucleon permutations. Each of the terms $V_{ijk}$ 
is also symmetric with respect to interchange of two of the three nucleons
(in our notation they are nucleons $i$ and $j$). When Jacobi vectors are 
constructed in the natural order the spatial components of basis states are 
chosen to be symmetric or antisymmetric with respect to the interchange of 
nucleons 1 and 2. Then it is convenient to use the $V_{123}$ component
of the force for the calculation of ME. The expression for this component 
is of the form \cite{carlsonetal83}  
\begin{eqnarray}
V_{123}=\bigl(A_{2\pi}\{x_{13},
x_{23}\}+B_{123}\{z_{13},z_{23}\}\bigr)
\{{\vec \tau}_1\cdot{\vec \tau}_3,{\vec \tau}_2\cdot{\vec \tau}_3\} \cr
+C_{2\pi}[x_{13},x_{23}]
[{\vec \tau}_1\cdot{\vec \tau}_3,{\vec \tau}_2\cdot{\vec \tau}_3]+
U_0\left(T_{13}T_{23}\right)^2.
\end{eqnarray}
Here $[\ldots\,,\,\ldots]$ and $\{\ldots\,,\,\ldots \}$ mean commutator and 
anticommutator, respectively, while 
\[x_{ij}={\vec \sigma}_i\cdot{\vec\sigma}_jY_{ij}+S_{ij}T_{ij},\qquad 
  z_{ij}={\vec \sigma}_i\cdot{\vec\sigma}_j+S_{ij},\]
where $Y_{ij}\equiv Y(r_{ij})$ and $T_{ij}\equiv T(r_{ij})$ are the regularized 
Yukawa and tensor functions, respectively. As customary $S_{ij}$ is the tensor 
operator, and $A_{2\pi}$, $C_{2\pi}$, and  $U_0$ are the strength constants
(for the Urbana models $C_{2\pi}=(1/4)A_{2\pi}$). The numerical values of 
$A_{2\pi}$ and $U_0$ that correspond to the Urbana IX model are listed in 
\cite{UrbIX}. The quantity $B_{ijk}\equiv B(r_{ik},r_{jk})$ is different from 
zero only in the Tucson-Melbourne model \cite{TM}.

In what follows we use the notation $\vec \sigma_i$ and $\vec \tau_i$ for 
nucleon spin and isospin operators. We use the following notation for tensor 
products of irreducible tensor operators
\begin{eqnarray*} 
(A_a\otimes B_b)_{\lambda\mu} & = & \sum_{\alpha+\beta=\mu}
   \bra a \alpha b \beta | \lambda\mu \ket A_{a\alpha}B_{b\beta}, 
  \qquad 
 (A_a\cdot B_a)=\sum_{\alpha}A_{a\alpha}B_{a,-\alpha}(-1)^{\alpha},
 \\ 
 (A_a\otimes B_b)_{\lambda\mu}^C & = & 
 \sum_{\alpha+\beta=\mu} \bra {a\alpha b\beta}|{\lambda\mu}\ket
                [A_{a\alpha},B_{b\beta}],
  \qquad 
 (A_a\otimes B_b)_{\lambda\mu}^A=\sum_{\alpha+\beta=\mu}
    \bra {a\alpha b\beta}|{\lambda\mu}\ket \{A_{a\alpha},B_{b\beta}\},
\end{eqnarray*} 
 and also the notation for irreducible spin tensors
\begin{equation} \label{spin}
  \Sigma_{ij}^{\lambda\mu} = (\vec\sigma_i \otimes \vec\sigma_j)_{\lambda\mu} 
  \,,\,\,
  \,\,\,\,\, \Sigma_{ij,k}^{\lambda, \Lambda M} = 
        \left(\vec\sigma_k \otimes 
              (\vec\sigma_i \otimes \vec\sigma_j)_{\lambda}
        \right)_{\Lambda M}.
\end{equation}
 
Use of spherical harmonics and their tensor products
as irreducible space 
tensors gives the expansion of $V_{123}$ in the form
\begin{eqnarray}
V_{123}=W_0+\left(Y_2(\hat{r}_{13})\cdot W_2^{TY}\right) +
            \left(Y_2(\hat{r}_{23})\cdot W_2^{YT}\right)+
            \sum_{\lambda=0}^3\left(Y_\lambda^{22}(\hat{r}_{13},\hat{r}_{23})
                  \cdot W_\lambda^{TT}\right),\label{f1}
\end{eqnarray} 
where $Y_{\lambda\mu}^{22}(\hat{r}_{13},\hat{r}_{23})
      =\left(Y_2(\hat{r}_{13})\otimes Y_2(\hat{r}_{23})\right)_{\lambda\mu}$, 
$\hat r_{ij}$ stands for the unit relative vector of nucleons $i$ and $j$ 
($\vec r_{ij} = \vec r_{i} -\vec r_{j}$). The quantities $W_0$, 
$W_{2\mu}^{TY}$, $W_{2\mu}^{YT}$ and $W_{\lambda\mu}^{TT}$ are irreducible 
tensors with respect to spin variables of the ranks 0, 2, and $\lambda$, 
respectively. One has
\begin{eqnarray}
W_0 & = & U_0(T_{13}T_{23})^2+(A_{2\pi}Y_{13}Y_{23}+B_{123})
 \{{\vec \tau}_1\cdot{\vec \tau}_3,{\vec \tau}_2\cdot{\vec \tau}_3\}
 \{{\vec \sigma}_1\cdot{\vec \sigma}_3,{\vec \sigma}_2\cdot{\vec \sigma}_3\}
 \nonumber\\ &&
 +C_{2\pi}Y_{13}Y_{23}[{\vec \tau}_1\cdot{\vec \tau}_3,{\vec \tau}_2
 \cdot{\vec \tau}_3]
 [{\vec \sigma}_1\cdot{\vec \sigma}_3,{\vec \sigma}_2\cdot{\vec \sigma}_3],
 \nonumber\\ 
W_{2\mu}^{TY}& = & \sqrt{\frac{24\pi}{5}}\left[(A_{2\pi}T_{13}Y_{23}+B_{123})
 \{{\vec \tau}_1\cdot{\vec \tau}_3,{\vec \tau}_2\cdot{\vec \tau}_3\}
 \{\Sigma^{2\mu}_{13},{\vec \sigma}_2\cdot{\vec \sigma}_3\}\right.
 \nonumber\\&&+
 \left.C_{2\pi}T_{13}Y_{23}
  [{\vec \tau}_1\cdot{\vec \tau}_3,{\vec \tau}_2\cdot{\vec \tau}_3]
  [\Sigma^{2\mu}_{13},{\vec \sigma}_2\cdot{\vec \sigma}_3]\right],
 \nonumber\\
W_{\lambda\mu}^{TT}&=&\frac{24\pi}{5}(-1)^\lambda
 \left[(A_{2\pi}T_{13}T_{23}+B_{123})
 \{{\vec \tau}_1\cdot{\vec \tau}_3,{\vec \tau}_2\cdot{\vec \tau}_3\}
 (\Sigma^2_{13}\otimes\Sigma^2_{23})_{\lambda\mu}^A\right.
 \nonumber\\&&
 \left.+C_{2\pi}T_{13}T_{23}
 [{\vec \tau}_1\cdot{\vec \tau}_3,{\vec \tau}_2\cdot{\vec \tau}_3]
 (\Sigma^2_{13}\otimes\Sigma^2_{23})_{\lambda\mu}^C\right],\label{c}
\end{eqnarray} 
and $W_{2\mu}^{YT}$ = (${\hat {12}}) W_{2\mu}^{TY}$.
The formulae above are obtained using the following expression for the
tensor operator 
\[S_{ij}\equiv3({\vec\sigma}_i\cdot{\hat r}_{ij})({\vec\sigma}_j\cdot{\hat r}_{ij})-
          ({\vec\sigma}_i\cdot{\vec\sigma}_j) 
        =\sqrt{\frac{24\pi}{5}}(Y_2({\hat r}_{ij})\cdot\Sigma^2_{ij})\]
and the recoupling
\[(Y_2({\hat r}_{13})\cdot\Sigma^2_{13})(Y_2({\hat r}_{23})\cdot\Sigma^2_{23})=
\sum_{\lambda=0}^3 (-1)^\lambda\bigl(Y_\lambda^{22}({\hat r}_{13},{\hat r}_{23})
\cdot(\Sigma^2_{13}\otimes\Sigma^2_{23})_\lambda\bigl)\]
(for $\lambda=4$ the quantity $(\Sigma^2_{13}\otimes\Sigma^2_{23})_{\lambda\mu}$ 
vanishes).

The ME of the scalar products entering (\ref{f1}) are expressed
in the standard way in terms of spatial reduced ME and spin--isospin ME reduced 
with respect to spin.
The latter ME may be calculated directly writing down all the quantities 
depending on spin and isospin
in terms of $\sigma_{i;0,\pm}$ and $\tau_{i;0,\pm}$. One  may also employ
the expansions over the
tensors of Eq. (\ref{spin}) form with subsequent use of the ME (\ref{s1}), 
(\ref{s2}). 
To obtain these expansions we perform recouplings, make use of the values of 
$\{\sigma_{3\mu},\sigma_{3\nu}\}$
and $[\sigma_{3\mu},\sigma_{3\nu}]$ \cite{Varshalovich} and also 
perform simple summations of Clebsch--Gordan coefficients
\cite{Varshalovich}. As a result we get the following
expression for the 3N force
\begin{eqnarray} \label{3NFv}
V_{123} & = & 4\sqrt{3} (\vec\tau_1 \otimes \vec\tau_2)_{0}
 \Big\{ (A_{2\pi} Y_{13}Y_{23}+B_{123})\sqrt{3}\Sigma_{12}^{0} 
       \cr & &  \hspace{2.1 cm}
 -\sqrt{\frac{24\pi}{5}}\left(1+(\hat{12})\right)(A_{2\pi} T_{13}Y_{23}+B_{123})
         (Y_2(\hat{r}_{13}) \cdot \Sigma_{12}^{2})
	          \cr & &  
 +5 \frac{24\pi}{5} (A_{2\pi} T_{13}T_{23}+B_{123})
     \sum_{\lambda=0}^3 \sqrt{3(2\lambda+1)}
        \ninej   {1}    {1}    {2} 
                 {1}    {1}    {2} 
              {\lambda} {0} {\lambda} 
             (Y_{\lambda}^{22}(\hat{r}_{13},\hat{r}_{23}) \cdot \Sigma_{12}^{\lambda}) 
           \Big\}
         \cr & & 
   +4 \sqrt{3} C_{2\pi}\left( \vec\tau_3 \otimes (\vec\tau_1 \otimes \vec\tau_2)_{1}
            \right)_0
 \Big\{
    2\sqrt{3}Y_{13}Y_{23}\Sigma_{12,3}^{1,0} 
         \cr & & \hspace{2.1 cm} 
  +\left(1+(\hat{12})\right)T_{13}Y_{23}\left(Y_2(\hat{r}_{13})\cdot(\Sigma_{12,3}^{1,2} 
       + \sqrt{3}\Sigma_{12,3}^{2,2}) \right)
         \cr & & 
       -10 \sqrt{3} T_{13}T_{23}
         \sum_{\Lambda=0}^3\Bigl(Y_{\Lambda}^{22}(\hat{r}_{13},\hat{r}_{23})\cdot 
	 \sum_\lambda (-1)^{\lambda+\Lambda}\sqrt{2\lambda+1}
        \ninej   {1}    {1}    {2} 
                 {1}    {1}    {2} 
              {\lambda} {1}  {\Lambda} 
	      \Sigma^{\lambda,\Lambda}_{12,3}\Bigl)
      \Big\}
         \cr & & 
         + U_0(T_{12}T_{13})^2.
\end{eqnarray}

Another multipole expansion of the force is obtained when the quantities 
\begin{equation}
  X_{ij}^{\lambda\mu} = (\hat r_{1i} \otimes \hat r_{1j} )_{\lambda\mu} \,,\,\,\,\,\,\,
  X_{ij,kl}^{(\lambda,\lambda')\Lambda M} 
     = \left( (\hat r_{1i} \otimes \hat r_{1j} )_{\lambda}
       \otimes (\hat r_{1k} \otimes \hat r_{1l} )_{\lambda'} \right)_{\Lambda M} \,.
\end{equation}
are used as spatial tensors  and, moreover,
an expansion of spin--isospin operators over the tensors of Eq. (\ref{spin}) form is performed. 
In this case we assume use of Jacobi vectors
constructed in the reversed order and for the calculation of ME 
we employ the component $V_{231}$ of the force. We rewrite it in the following form:
\begin{eqnarray}
V_{231}& = & 
   A_{2\pi} ( 2 {\vec\tau}_2\cdot{\vec\tau}_3 )   \big( 
   F_{TT} (\vec\sigma_2\cdot\hat r_{12})
          (\vec\sigma_3\cdot\hat r_{13})(\hat r_{12} \cdot \hat r_{13}) 
    + F_{YY} ({\vec\sigma}_2\cdot{\vec\sigma}_3)
   \cr & & \hspace {1 cm} +  
   F_{TY} (\vec\sigma_2\cdot\hat r_{12})(\vec\sigma_3\cdot\hat r_{12})
   + F_{YT} (\vec\sigma_2\cdot\hat r_{13})(\vec\sigma_3\cdot\hat r_{13})
   \big) \cr & & -
   C_{2\pi} ( 2 \vec\tau_1\cdot \vec\tau_2 \times \vec\tau_3 )   \big( 
   F_{TT} (\vec\sigma_2\cdot\hat r_{12})(\vec\sigma_3\cdot\hat r_{13})
          (\vec\sigma_1\cdot \hat r_{12} \times \hat r_{13}) 
    + F_{YY} (\vec\sigma_1\cdot\vec\sigma_2\times\vec\sigma_3)
   \cr & & \hspace {1 cm} +  
   F_{TY} (\vec\sigma_2\cdot\hat r_{12})
          (\vec\sigma_3\cdot\vec\sigma_1\times\hat r_{12})
   + F_{YT} (\vec\sigma_3\cdot\hat r_{13})
          (\vec\sigma_1\cdot\vec\sigma_2\times\hat r_{13})
   \big) \cr & + &
   B_{231} ( 2 {\vec\tau}_2\cdot{\vec\tau}_3 )    
   18 (\vec\sigma_2\cdot\hat r_{12})
                (\vec\sigma_3\cdot\hat r_{13})(\hat r_{12} \cdot \hat r_{13}) 
   + U_0(T_{12}T_{13})^2 
\end{eqnarray}	
with   $F_{TT}= 18 T_{12}T_{13}$ and
\begin{equation}
  F_{YY}=  2 (Y_{12}-T_{12})(Y_{13}-T_{13}) \,,\,\,\,
  F_{TY}=  6 T_{12}(Y_{13}-T_{13}) \,,\,\,\,
  F_{YT}=  6 (Y_{12}-T_{12})T_{13} \,.
\end{equation}

After a recoupling of the various operators one obtains
\begin{eqnarray} \label{3NF}
V_{231} & = & -2 \sqrt{3} A_{2\pi} (\vec\tau_2 \otimes \vec\tau_3)_{0}
 \Big\{
    \Sigma_{23}^{0} \cdot [ -\sqrt{3}(F_{TT} + 18 \frac{B_{231}}{A_{2\pi}})X_{23}^0 X_{23}^0
                                 -\frac{1}{\sqrt{3}}(3F_{YY}+F_{TY}+F_{YT})]
         \cr & &  \hspace{3.1 cm}
 + \Sigma_{23}^{1} \cdot [\sqrt{3} (F_{TT} + 18 \frac{B_{231}}{A_{2\pi}})X_{23}^0 X_{23}^1 ]
         \cr & &  \hspace{3.1 cm}
 + \Sigma_{23}^{2} \cdot [-\sqrt{3} (F_{TT} + 18 \frac{B_{231}}{A_{2\pi}} )X_{23}^0 X_{23}^2
                           +   (F_{TY} X_{22}^2 +F_{YT} X_{33}^2) ]
           \Big\}
         \cr & & 
   -4 \sqrt{3} C_{2\pi}\left( \vec\tau_1 \otimes (\vec\tau_2 \otimes \vec\tau_3)_{1}
            \right)_0
 \Big\{
    \Sigma_{23,1}^{1,0} \cdot \sqrt{3} [ \sixj{1}{1}{1}{1}{0}{1} (F_{TY}+F_{YT})
                              - F_{YY} 
                              - \frac{F_{TT}}{\sqrt{3}} X_{23,23}^{(1,1)0}]
         \cr & & \hspace{4.3 cm} 
  +  \Sigma_{23,1}^{0,1} \cdot [F_{TT} X_{23,23}^{(1,0)1} ]
  +  \Sigma_{23,1}^{2,1} \cdot [F_{TT} X_{23,23}^{(1,2)1} ]
         \cr & & \hspace{4.3 cm} 
  +  \Sigma_{23,1}^{1,2} \cdot [-3 \sixj{1}{1}{1}{1}{2}{1} 
                                              (F_{TY}X_{22}^2+F_{YT}X_{33}^2)
                               - F_{TT} X_{23,23}^{(1,1)2}]
         \cr & & \hspace{4.3 cm} 
  +   \Sigma_{23,1}^{2,2} \cdot [ 
         \sqrt{15} \sixj{1}{1}{2}{1}{2}{1} (F_{TY}X_{22}^2-F_{YT}X_{33}^2) 
                                - F_{TT} X_{23,23}^{(1,2)2}]
         \cr & & \hspace{4.3 cm} 
  +  \Sigma_{23,1}^{2,3} \cdot [ F_{TT} X_{23,23}^{(1,2)3} ] 
         \Big\} + U_0(T_{12}T_{13})^2.
\end{eqnarray}

For the spin operators of Eq.~(\ref{spin}) one finds the following reduced 
matrix elements 
\begin{eqnarray}
\label{s1}\lefteqn{
     \langle (s_1;(s_2;s_3)S_{23})S_{123} || 
     ( \vec\sigma_2 \otimes \vec\sigma_3)_{\lambda}
     || (s_1;(s_2;s_3)S'_{23})S'_{123} \rangle}  \cr
 & = & 
     6 (-)^{1/2+\lambda+S'_{23}+S_{123}}
     \sqrt{(2 S_{123} + 1)(2 S'_{123} + 1)}
     \sixj{S_{23}}{S_{123}}{\frac{1}{2}}{S'_{123}}{S'_{23}}{\lambda}
     \cr & &
    \times \sqrt{(2 S_{23} + 1)(2 S'_{23} + 1)(2 \lambda + 1)}
     \ninej   {\frac{1}{2}}  {\frac{1}{2}}  {S_{23}}
              {\frac{1}{2}}  {\frac{1}{2}}  {S'_{23}}
                   {1}           {1}        {\lambda} \,,
\end{eqnarray} 
\begin{eqnarray}
\label{s2}\lefteqn{
     \langle (s_1;(s_2;s_3)S_{23})S_{123} || 
     (\vec\sigma_1 \otimes 
     (\vec\sigma_2 \otimes \vec\sigma_3)_{\lambda})_{\Lambda}
     || (s_1;(s_2;s_3)S'_{23})S'_{123} \rangle } \cr
 & = & 
     6^{\frac{3}{2}} 
     \sqrt{(2 S_{123} + 1)(2 S'_{123} + 1)(2 \Lambda + 1)}
     \ninej   {\frac{1}{2}}  {S_{23}}   {S_{123}} 
              {\frac{1}{2}}  {S'_{23}}	{S'_{123}}
                   {1}       {\lambda}	{\Lambda}
     \cr & &
     \times\sqrt{(2 S_{23} + 1)(2 S'_{23} + 1)(2 \lambda + 1)}
     \ninej   {\frac{1}{2}}  {\frac{1}{2}}  {S_{23}}
              {\frac{1}{2}}  {\frac{1}{2}}  {S'_{23}}
                   {1}           {1}        {\lambda} \,.
\end{eqnarray} 
The isospin operators of Eqs.~(\ref{3NF}) and (\ref{3NFv}) lead to corresponding expressions
for the reduced matrix elements.

For the calculation of the spatial matrix elements one has to perform a
six--dimensional integration. For this integration we use the same technique as 
first used in \cite{ELOT} and explicitly described in \cite{Efros}. 
For the Jacobi vectors (see section III)
\begin{equation}
     {\vec \eta}_{A-1}  = \rho^{[3]}_{int} \,\sin 
      \theta^{[3]}_{int} \, {\hat \eta}_{A-1} \,,\,\,\,\,\,\,
     \,\,\,\, \, {\vec \eta}_{A-2} = \rho^{[3]}_{int}  \,\cos 
      \theta^{[3]}_{int} \, {\hat \eta}_{A-2} 
\end{equation}
one chooses a special reference frame in which ${\hat \eta}_{A-1}$ turns into 
a vector ${\hat \zeta}_{A-1} = \hat z$ 
and ${\hat \eta}_{A-2}$ turns into a vector ${\hat \zeta}_{A-2}$ that 
lies in the ($x$-$z$) plane. 
Then one replaces the 
angles $\theta^{[3]}_{int}$, ${\hat \eta}_{A-1}$, and ${\hat \eta}_{A-2}$ by
\begin{equation}
  x = cos (2 \theta^{[3]}_{int})\,, \hspace{2cm} 
  y = {\hat \eta}_{A-1} \cdot {\hat \eta}_{A-2} \,,
\end{equation}
and the three Euler angles $\omega$. This leads to the following volume element
\begin{equation}
   d{\vec \eta}_{A-1} d{\vec \eta}_{A-2} = (\rho^{[3]}_{int})^5 
                 d\rho^{[3]}_{int} \frac{1}{8}\sqrt{1-x^2}
                 dx d{\hat \eta}_{A-1} d{\hat \eta}_{A-2}
             = (\rho^{[3]}_{int})^5 d\rho^{[3]}_{int}
                \frac{1}{8}\sqrt{1-x^2} dx dy 
               d\omega \,.
\end{equation}
With this parametrization one obtains for the three--body HH functions
\begin{eqnarray}
  Y_{K L M}^{\ell_1 \ell_2}(\Omega) & = & 
   \psi_{K}^{\ell_1 \ell_2}(x) 
   [Y_{\ell_1 m_1}({\hat \eta}_{A-1}) \otimes 
    Y_{\ell_2 m_2}({\hat \eta}_{A-2})]_{L M}
\cr & = &
   \psi_{K}^{\ell_1 \ell_2}(x) 
   \sum_{M'} D^{(L)}_{M' M}(\omega)
   [Y_{\ell_1 m_1}({\hat \zeta}_{A-1}) \otimes 
    Y_{\ell_2 m_2}({\hat \zeta}_{A-2}]_{L M'}
\cr & = &
   \sum_{M'} D^{(L)}_{M' M}(\omega)
   Y_{K L M'}^{\ell_1 \ell_2}(x,y) \,.
\end{eqnarray}
The five--dimensional integration in $dx dy d\omega$ can be reduced to a 
two--dimensional one. In fact after some algebra one finds for a coordinate 
space operator $V_{\lambda\mu}(\rho^{[3]}_{int},\Omega^{[3]}_{int})$ 
of rank $\lambda$ and projection
$\mu$ the following reduced matrix element 
\begin{eqnarray}
  \langle K L \ell_1 \ell_2 || 
     V_{\lambda}(\rho^{[3]}_{int},\Omega^{[3]}_{int}) 
                                ||  K' L' \ell'_1 \ell'_2 \rangle
  & = & \sum_{M M'\mu} (-)^{L+M}\threej{L}{\lambda}{L'}{-M}{\mu}{M'}
 \cr & & \hspace{-4cm}
 \times \pi^2 \int dx \sqrt{1-x^2} dy Y_{K L M}^{\;\ell_1 \ell_2 * }(x,y)
  V_{\lambda\mu}(\rho^{[3]}_{int},x,y) Y_{K' L' M'}^{\ell'_1 \ell'_2}(x,y) \,.
\end{eqnarray}
\subsection*{Acknowledgments}

The work of N.B. was supported by the {\bf ISRAEL SCIENCE FOUNDATION}
(grant no. 202/02), V.D.E. acknowledges the support of the Russian Ministry
of Industry and Science, grant NS--1885.2003.2.


\vfill\eject

\begin{table}
\caption{$^3$H ground state properties with AV18+UrbIX (no electromagnetic
interaction, no isospin mixing) for binding energy BE
and probabilities of total orbital angular momentum components in \%}
\renewcommand{\arraystretch}{1.1}
\begin{tabular}{ccc}
  \   &\ this work \  &\ \  CHH \cite{ELOT2} \\
\hline

BE [MeV]                 &  8.508    &  8.47   \\
S-wave                   & 89.504    & 89.55   \\
S$^\prime$-wave          &  1.061    &  1.05   \\
P-wave                   &  0.135    &  0.13   \\
D-wave                   &  9.300    &  9.27   \\
\end{tabular}
\end{table}

\begin{table}
\caption{Same as Table~I but with inclusion of electromagnetic
interaction (no isospin mixing) }
\renewcommand{\arraystretch}{1.1}
\begin{tabular}{cccc}
  \   &\ this work \  &\ \  Bochum \cite{BoPi} \ \ &\ \ Pisa \cite{BoPi}\\
\hline

BE [MeV]                 &  8.468  &  8.470  &  8.474  \\
S-wave                   & 89.516  & 89.512  & 89.509  \\
S$^\prime$-wave          & 1.059   &  1.051  &  1.055  \\
P-wave                   & 0.135   &  0.135  &  0.135  \\
D-wave                   & 9.291   &  9.302  &  9.301  \\
\end{tabular}
\end{table}

\begin{table}
\caption{Same as Table~II but for $^3$He}
\renewcommand{\arraystretch}{1.1}
\begin{tabular}{cccc}
  \   &\ this work \  &\ \  Bochum \cite{BoPi}\ \ &\ \ Pisa \cite{BoPi} \\
\hline

BE [MeV]                 &  7.736    &  7.738   &  7.742  \\
S-wave                   & 89.385    & 89.391   & 89.378  \\
S$^\prime$-wave          &  1.246    &  1.229   & 1.242   \\
P-wave                   &  0.132    &  0.132   & 0.131   \\
D-wave                   &  9.238    &  9.248   & 9.249   \\
\end{tabular}
\end{table}

\vfill\eject

\begin{figure}
\rotatebox{0}{
\resizebox{12cm}{!}{
\includegraphics{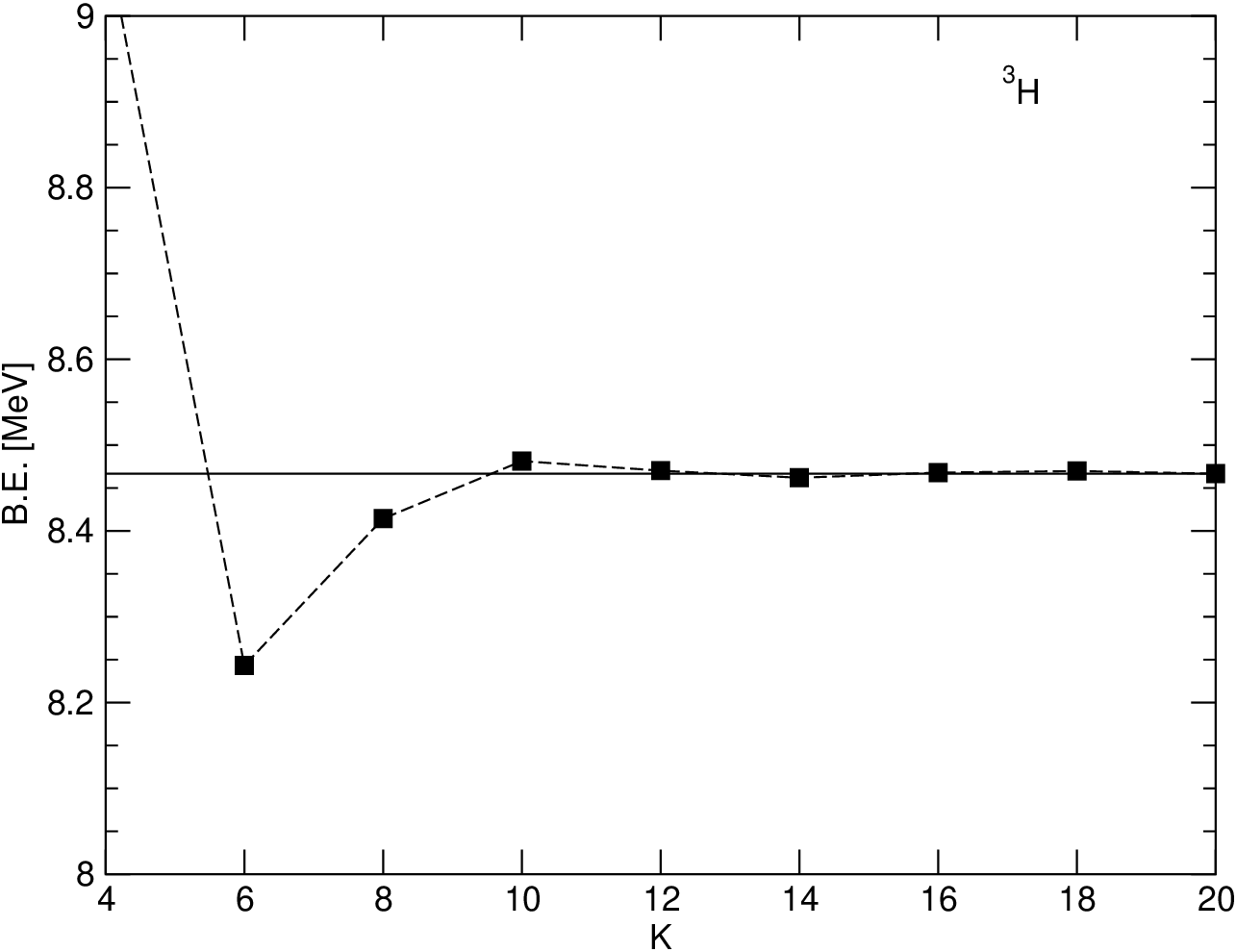}}}
\caption{Triton binding energy with AV18+UrbIX (no isospin mixing)
as function of grand angular momentum quantum number $K$.
\label{fig1}}
\end{figure}

\begin{figure}
\rotatebox{0}{
\resizebox{12cm}{!}{
\includegraphics{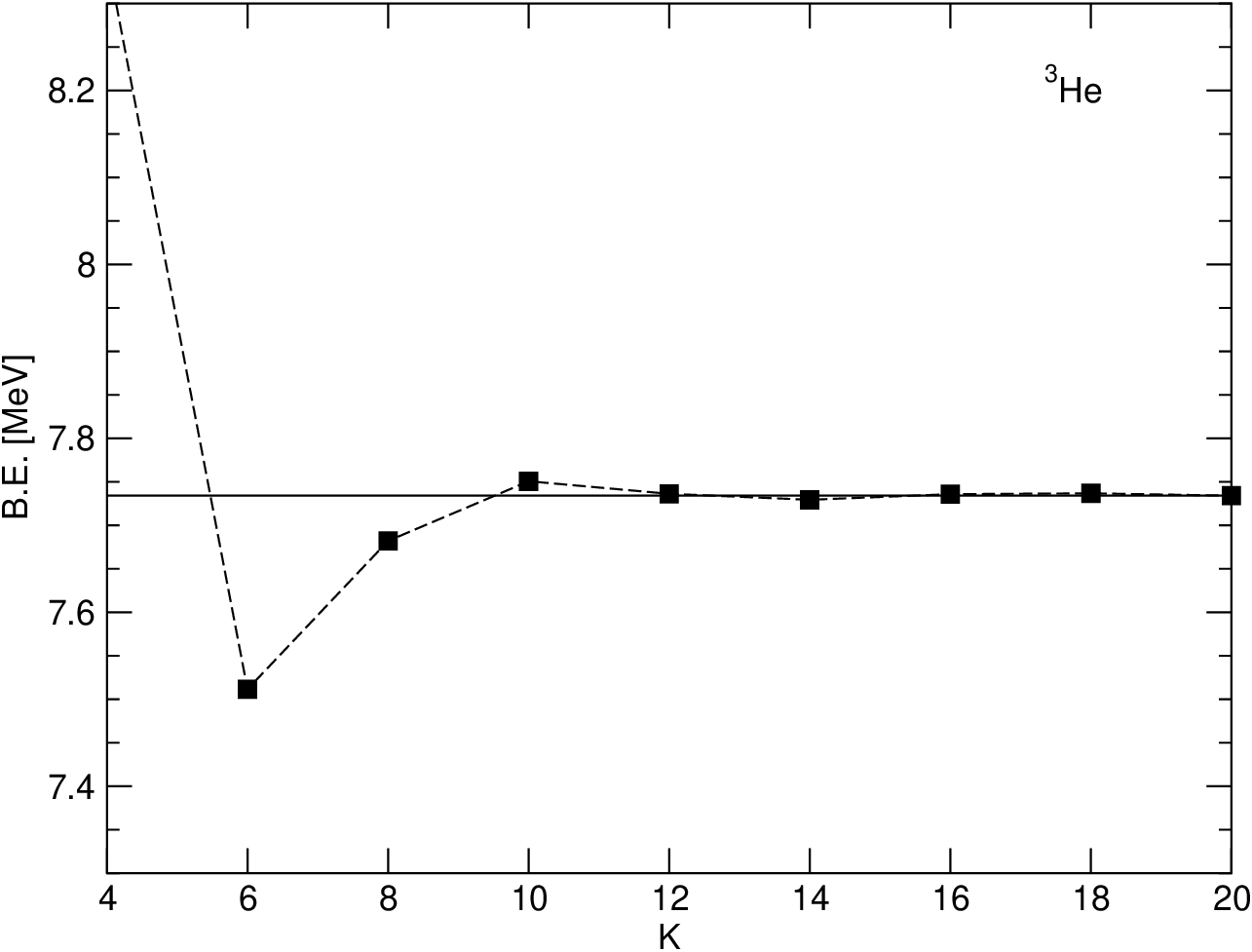}}}
\caption{Same as Fig.~1 but for $^3$He.
\label{fig2}}
\end{figure}

\begin{figure}
\rotatebox{0}{
\resizebox{12cm}{!}{
\includegraphics{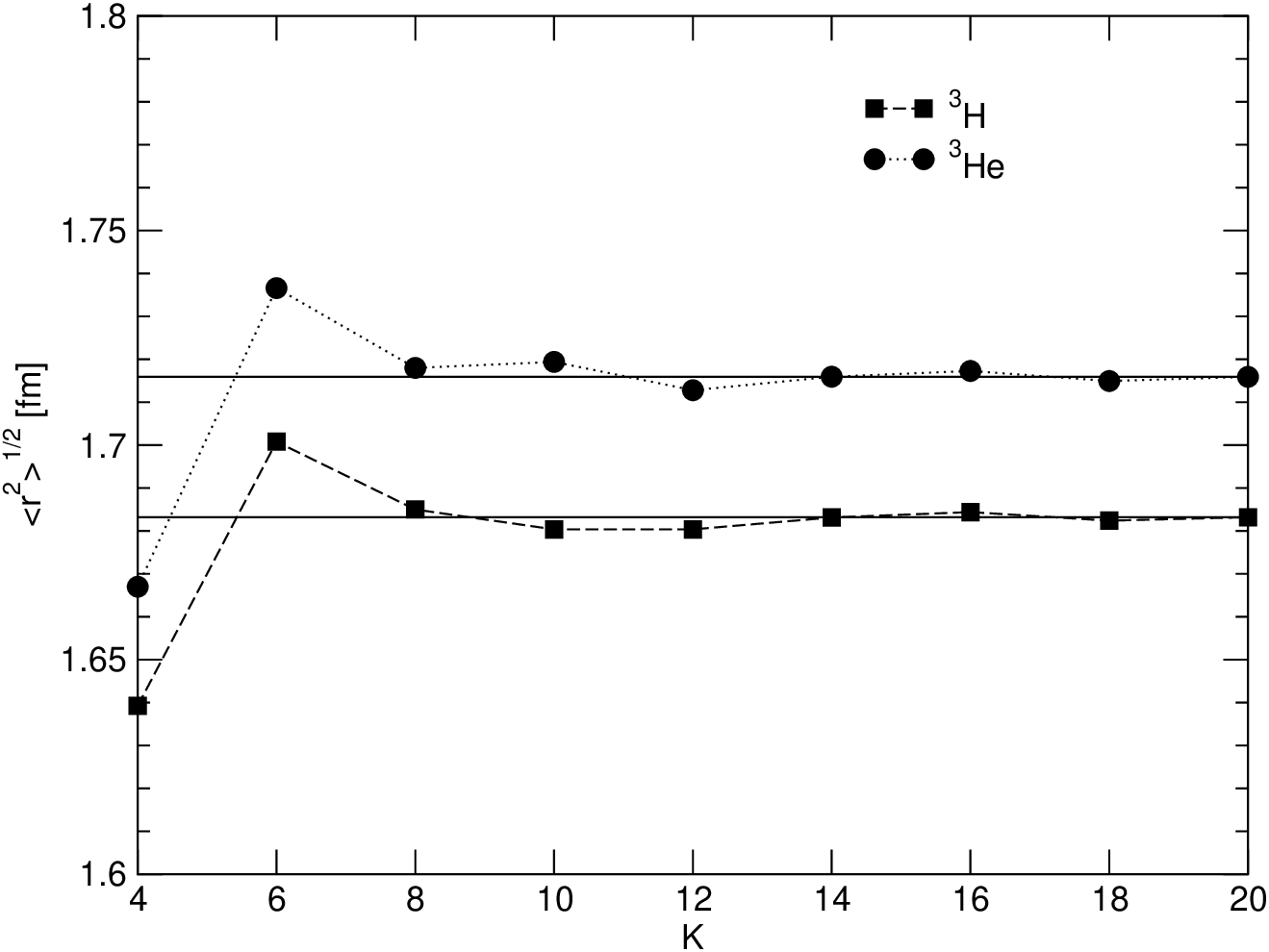}}}
\caption{$^3$H and $^3$He matter radii with AV18+UrbIX (no isospin mixing)
as function of grand angular momentum quantum number $K$.
\label{fig3}}
\end{figure}

\end{document}